# How Fragmentation Can Undermine the Public Health Response to COVID-19

Dr. Andrew Tzer-Yeu Chen
Koi Tū: The Centre for Informed Futures, University of Auckland, New Zealand
andrew.chen@auckland.ac.nz

Responses to COVID-19 have largely been led by local, national, and international public health agencies, who have activated their pandemic plans and opened the epidemiological toolkit of modelling, testing, isolation and movement restrictions, surveillance, and contact tracing. In the contemporary tech-heavy world, many assumed that the common manual process of human investigators and phone calls could or should be replaced by digital solutions. But it's not as simple as "add more technology" – the complex way in which users and societies interact with the technology has significant impacts on effectiveness. When efforts are not well co-ordinated, fragmentation in system design and user experience can negatively impact the public health response. This article briefly covers the journey of how contact tracing registers and digital diaries evolved in New Zealand during the COVID-19 pandemic, the initial poor outcomes caused by the lack of central co-ordination, and the later improvement.

## What are Contact Tracing Registers and Digital Diaries?

New Zealand epidemiologist Dr. Ayesha Verrall notes that "rapid case detection and contact tracing, combined with other basic public health measures, has over 90% efficacy against COVID-19 at the population level, making it as effective as many vaccines." [1] Contact tracing involves identifying people who have been in contact with an infected person, and therefore who may have been unknowingly exposed to the infectious disease. By identifying the contacts and rapidly isolating and testing those individuals, the chains of transmission in the community are cut off, limiting the spread of the disease. Importantly, contact tracers need to find potential contacts who are unknown to the infected person, and can only do this by tracing the movement of the person to find others who may have overlapped in time and place. Contact tracing is not a new process – it has been used since the 1900s.

With the pervasive nature of digital technologies, there has been a lot of discussion globally around digital contact tracing solutions, particularly around Bluetooth-enabled smartphone apps (including the Apple/Google protocol) and wearable devices [2,3]. Proponents of the technology offer the promise that these solutions can achieve better completeness (finding more contacts of cases, especially where the identities are not known to the case) and speed (finding contacts and testing/isolating them faster). However, in the absence of a validated, effective digital contact tracing solution initially, a number of governments opted for simpler, lower-tech methods of collecting data about people's movements.

The contact tracing register (or visitor/customer/check-in log) has been deployed around the world – individuals are asked to provide their personal details at businesses and other places of interest, so that if a venue is identified as potential exposure site, then the register can be provided to contact tracers to quickly find people who were there at the relevant time. Digital diaries have also been introduced to help people keep track of their own movements to support their recollection if they get interviewed by a contact tracer – the distinction being that instead of the venue or the government holding the records, the individual themselves maintain and control their logs.

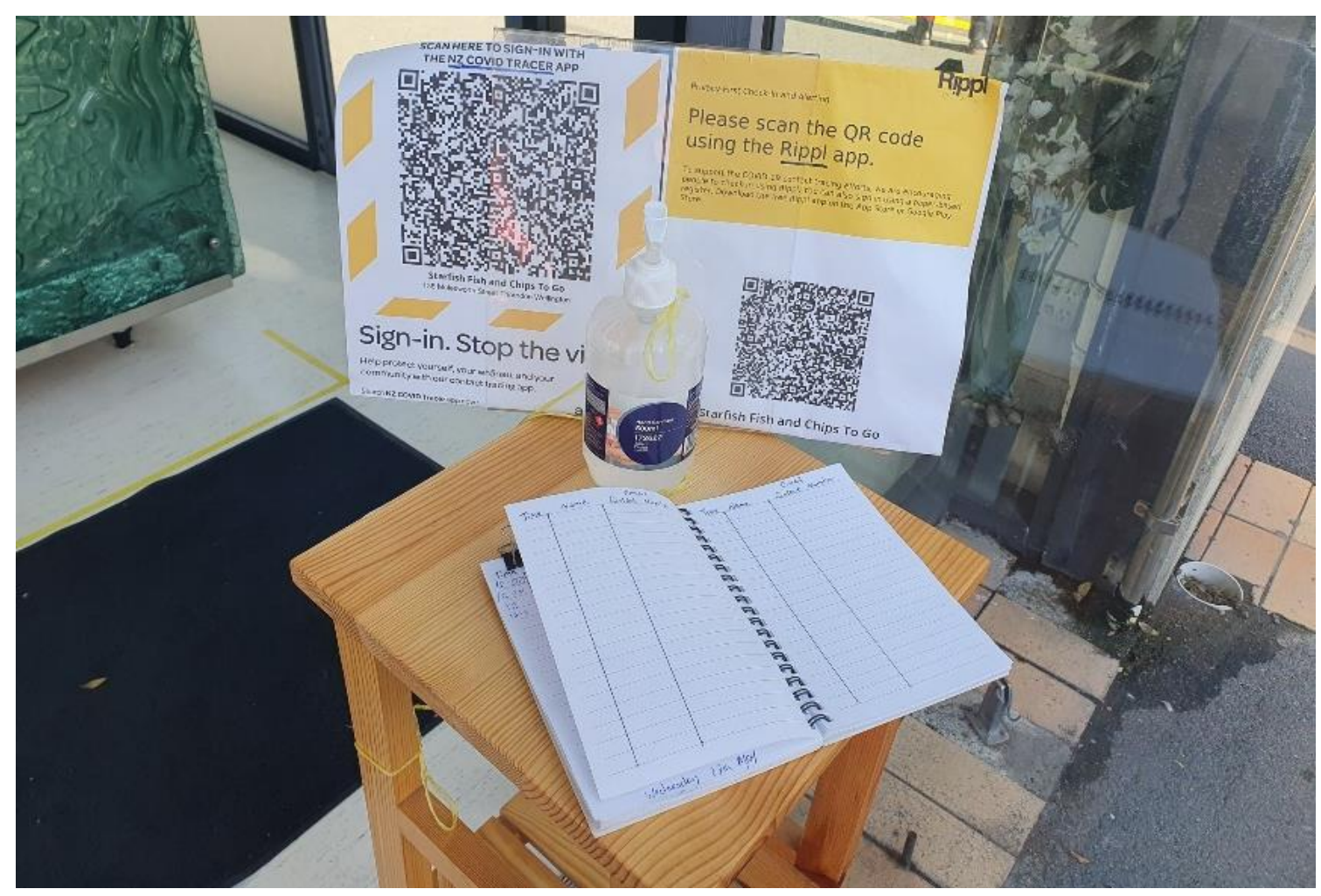

Figure 1. A pen-and-paper contact tracing book along with two QR codes for digital diaries in Wellington, New Zealand.

## Too Many Solutions

In New Zealand, one of the few countries that pursued an elimination strategy towards managing COVID-19, a strict Level 4 lockdown was implemented across the country on March 25. Most people (with the exception of essential services) stayed at home, and the high level of compliance meant that within two weeks the number of active cases began to fall. After four weeks, on April 27, some restrictions were eased at Level 3 (particularly around schools and takeaway food), and then most restrictions were lifted at Level 2 on May 13 (with the exception of gathering limits). As the country moved into Level 3, the government introduced a requirement under the Public Health Response Order for all businesses to maintain contact tracing registers. These registers required visitors and customers to provide their entry/exit times, name, address, and contact details to the business in case they are needed for contact tracing purposes. The government provided a template for businesses to print and use.

A number of criticisms were levelled at the pen-and-paper contact tracing registers that most businesses used initially. First, personal information on a piece of paper visible to all customers created a privacy risk. This led to real privacy breaches, such as a female customer being harassed by a male restaurant worker after he took her details from a contact tracing register. There were also some concerns about "dirty pen" risks (if everyone is using the same pen, could that become a vector for virus transmission?), usability (can people be bothered providing their details at every business they go to?), and validity (could people provide false details?).

Private software developers took the initiative to come up with better solutions. They reasoned that most people have smartphones (NZ has 80-85% smartphone penetration), and that using digital tools would mitigate or resolve some of the risks associated with pen-and-paper approaches. Within a week, there were over 30 tools available, almost all using QR codes, with a variety of system architectures and user flows. Some QR codes directed the user to a URL (thus requiring a mobile internet connection), others required a specific standalone app to interpret the code. Some stored the data on a central server owned and controlled by the developer, others stored the data on the phone for the user's reference only in a decentralized way (i.e. the "digital diary" approach). Some collected only a name and contact email address, others also asked for phone numbers and residential addresses – it was unclear what would be genuinely necessary for contact tracers to find people. Some tools were offered for free, others required businesses to pay a monthly fee, and two City Councils negotiated bulk licenses of one product for their cities. Some providers had developed full privacy policies, others said that speed-to-deployment was more important. Unfortunately, there was duplication of effort, and many developers found themselves re-inventing the wheel.

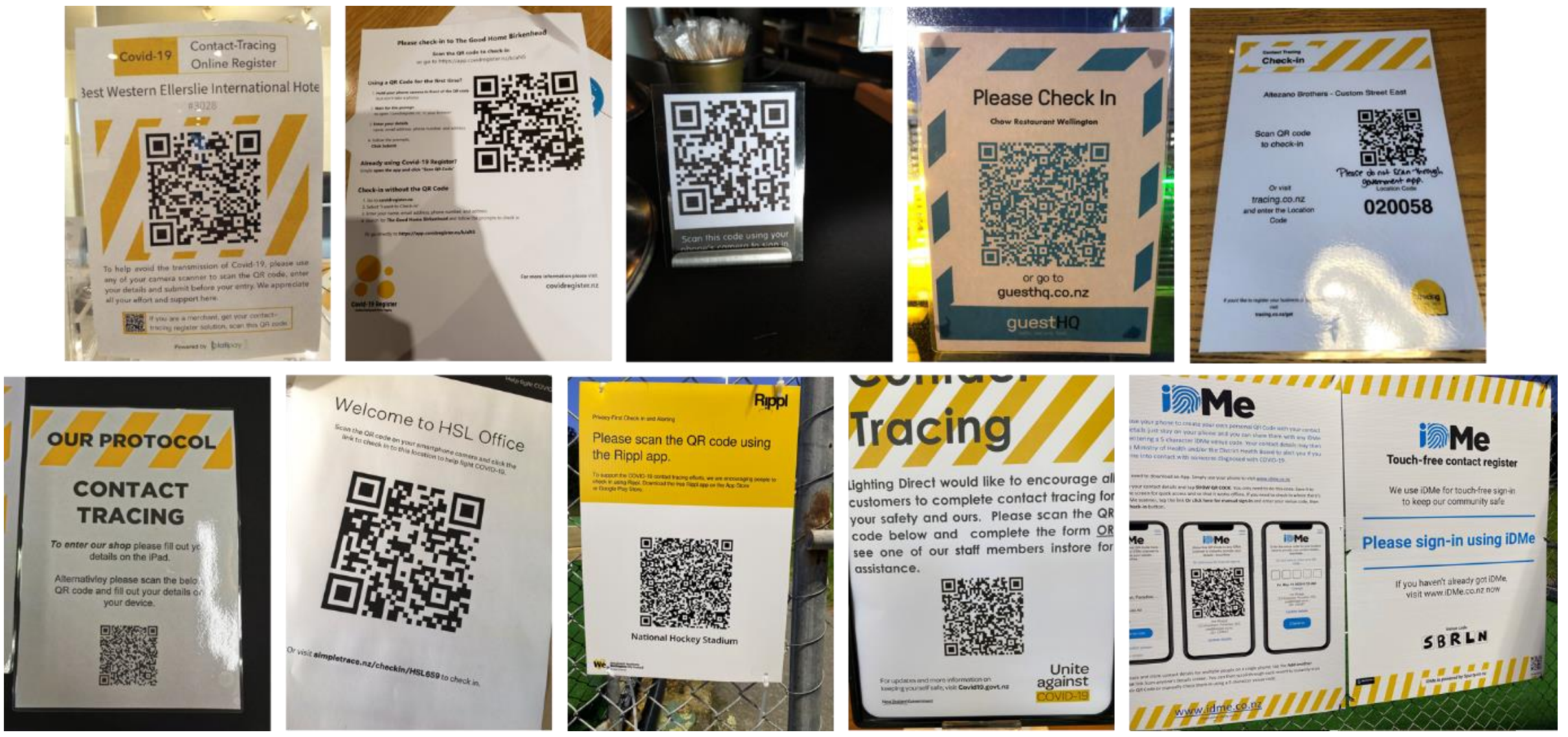

Figure 2. Photos of various QR codes from different providers in New Zealand, crowdsourced by the author from Twitter.

Almost every business soon adopted a QR code from one of these private providers, but with very little co-ordination or information around which systems were trustworthy or superior, chaos ensued. A lack of familiarity with how QR codes work amongst the public also led to significant confusion, with people getting frustrated when some QR codes worked and others didn't. This was not helped by a number of the QR code posters using similar branding, such as the yellow diagonal stripes that were used in government messaging about COVID-19, and some posters even using government logos to make their posters look more official. Most of the systems had centralized approaches, which also led to concerns about security and potential unauthorized reuse of data held by private corporations. Some businesses simply created web forms and added clauses in their Privacy Policies that allowed them to re-use the collected data for marketing purposes, attracting a stern message from the Privacy Commissioner. An interesting counterargument was that since there were so many different tools, the data was fragmented between different providers and therefore no single company held too much power.

This fragmented approach also meant that there was insufficient consideration for the needs of marginalized people. Posters were sometimes placed in positions that were inaccessible for disabled populations. Some businesses removed pen-and-paper registers entirely, making it impossible for participation by the digitally excluded – people without smartphones, or without the skills to effectively use the smartphone, or without expensive mobile internet data. There was also some confusion about whether or not digital diary approaches (with the data staying on the device) complied with the regulatory requirement for businesses to maintain registers.

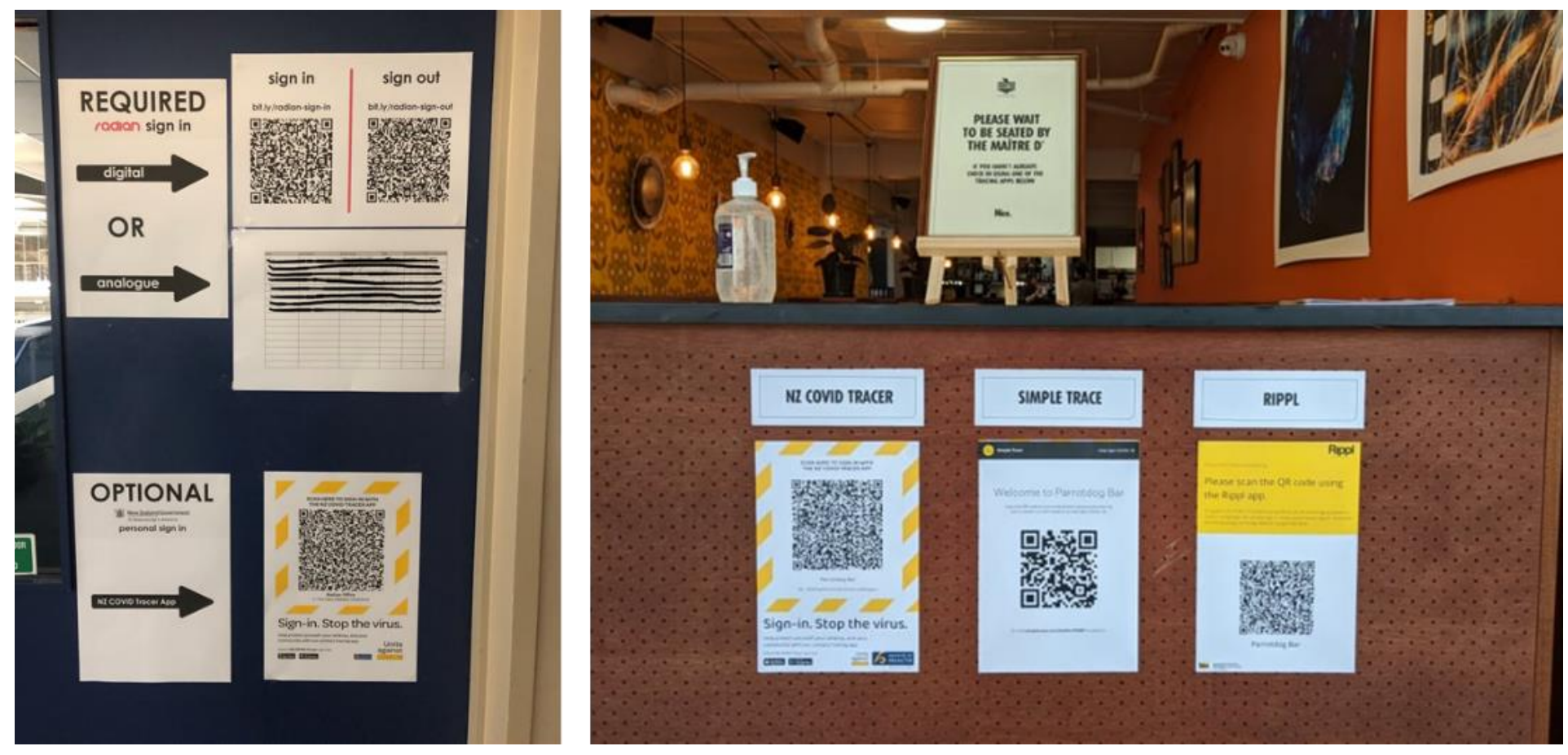

Figure 3. Businesses attempting to provide clearer instructions on which QR codes to use.

## The Government Steps In

On May 20, a week into Level 2, the Ministry of Health launched the [NZ COVID Tracer app](). This was (and is) also a QR code-based system with a "digital diary" approach. The app was accompanied by its own QR code standard, which contained a unique Global Location Number for each business, and therefore could be scanned without requiring an internet connection. Data about check-ins (where people had been at what time) stayed on the device, and the user can choose to release that information to a human contact tracer if identified as a close contact of a known case. The app also allowed individuals to provide up-to-date contact details to the Ministry of Health, which would help contact tracers find them more quickly if necessary. However, the government made it clear that the app would augment, not replace, manual contact tracing, and humans would clearly remain in-the-loop.

Unfortunately, QR codes were everywhere by this stage. Some businesses tried to provide multiple options (as shown in Figure 2) with clearer instructions, but this was limited. The two loudest complaints about the government app were that a) it wasn't compatible with older devices (requiring at least Android 7.0 or iOS 12 at launch) and that b) it didn't recognize most of the QR codes available. The government app wasn't designed to work with the other QR codes (which from a technical perspective might seem obvious, but for non-technical folks was bewildering). Displaying the government QR code was not mandatory, so many businesses didn't even have it as an option.

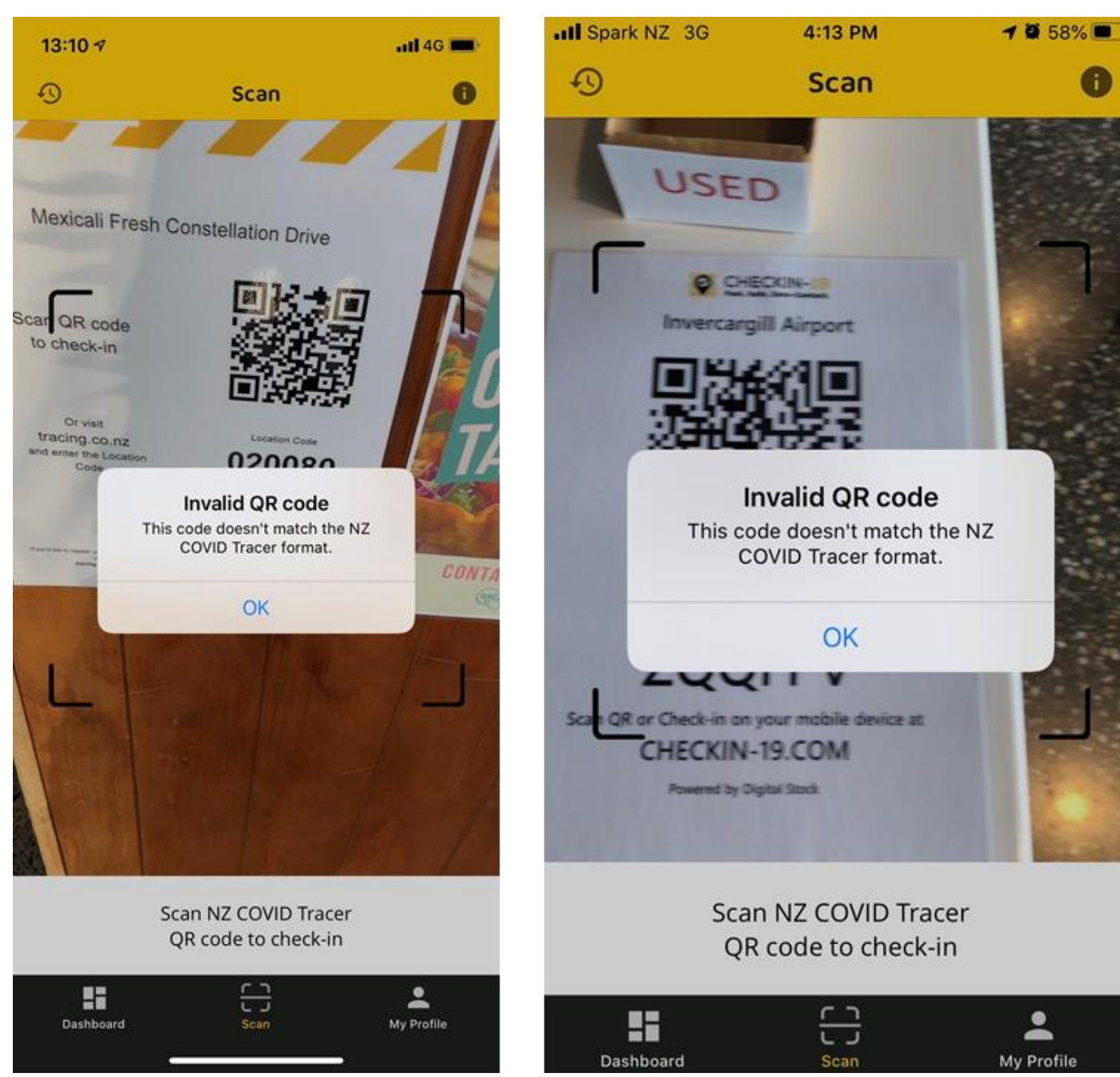

Figure 4. Screenshots from users complaining that the NZ COVID Tracer app wasn't recognizing the QR code, when they were in fact scanning QR codes from other providers.

This fragmentation harmed the uptake of the government app because people felt that existing tools served the same purpose. Within a week of launch, about 380,000 users were registered, equivalent to approximately 10% of the adult population of four million people. Registrations plateaued quickly, and a month later sat around 570,000. Meanwhile, the number of QR codes being scanned each day was counted through web analytics events, slowly ramping up initially as businesses started printing and displaying government QR codes, settling around 50,000 scans per day in early June. Given the size of the population, this was clearly not enough activity to give us confidence that the data from the app would be useful in the event of a further outbreak. However, the government app was the only one that reported usage statistics, so we don't have data about how widely other tools might have been used.

It turned out that the government app had actually been in development (with a private sector partner) for at least a month. The specific reasons for why the app was released late have not been made clear yet, although it should be noted that the government has a higher onus to "do things correctly" and needed to prepare a full privacy impact assessment, undergo an independent security audit, have the app checked by the government cybersecurity bureau, and complete other steps that weren't required for private developers.

By July, it appeared that New Zealand had the pandemic under control, leading to 102 days in a row without any community cases (excluding those caught and quarantined at the border) of COVID-19 detected. The Prime Minister moved the country to Level 1, lifting almost all restrictions except for border controls. The resulting sense of complacency was manifested by daily scan counts dropping to 10,000 in early July, and reports that some businesses were taking their posters down because they felt the QR codes were no longer necessary.

An improvement to the app in June introduced Exposure Notification functionality – human contact tracers can identify and securely broadcast a place and time where an active case had been, and then the app would check that against the check-in logs on the device and notify the user if an overlap was found. In late July, a further improvement was made to allow users to add manual entries, mostly to account for venues that did not have a government QR code. However, while there were minor upticks in registration and usage after these releases, the activity level still remained very low.

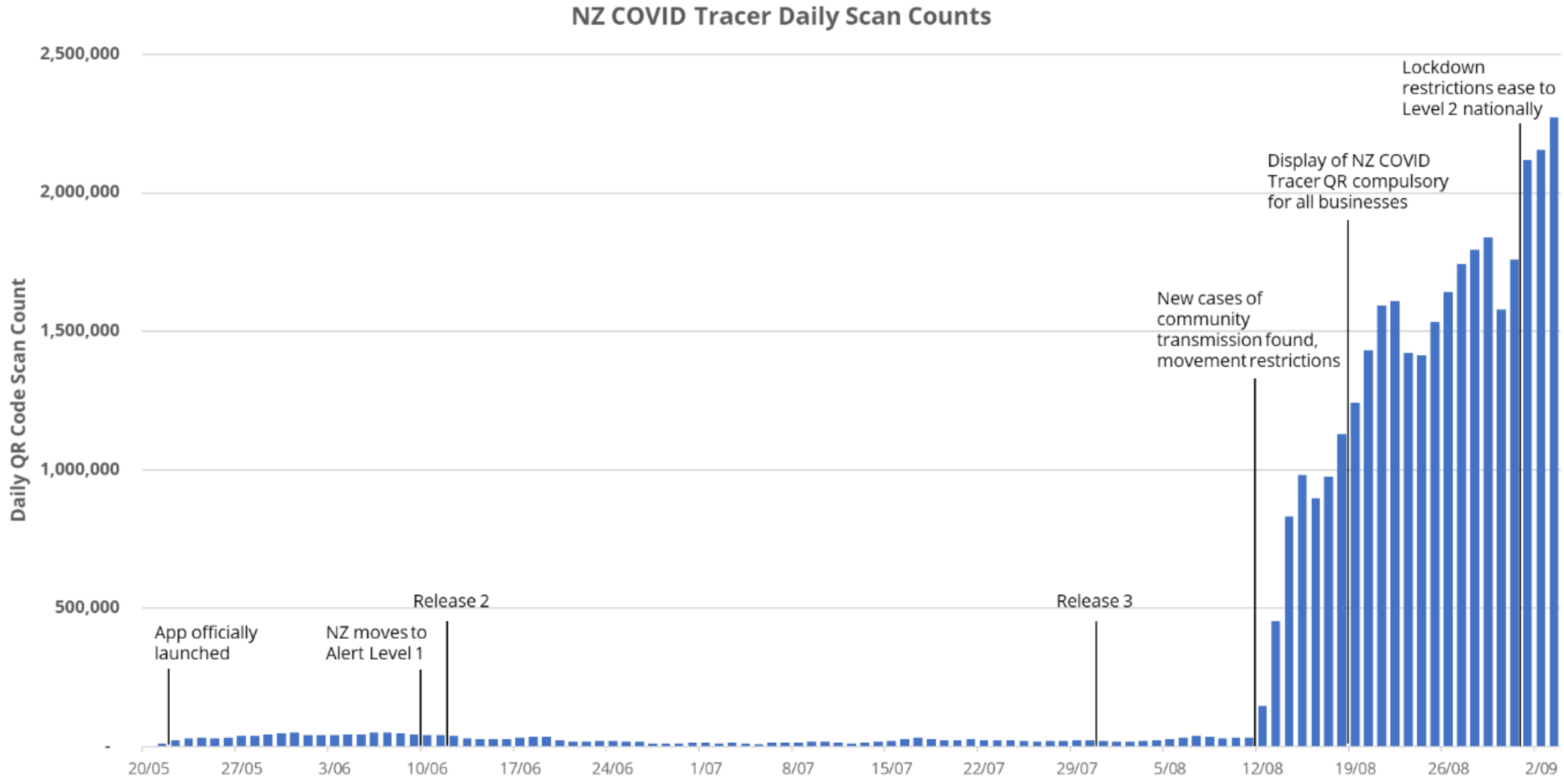

Figure 5. Daily scan counts from the NZ COVID Tracer app, overlaid with significant events. Data sourced from the NZ Ministry of Health.

## Consolidation Drives Uptake

On August 11, the Prime Minister announced that four cases of community transmission found in Auckland (the largest city in New Zealand) had no links to overseas travel, so it was highly likely that there were other undetected cases in the community. Given the potential risk of a rapidly growing second wave, the government implemented stricter movement restrictions around Auckland and heightened the alert level nationally. Displaying a NZ COVID Tracer QR code in a prominent place would became mandatory for all businesses by the following week. While still be optional for individuals to scan, at least the codes were available for people to scan if they wanted to. This decision wasn't without precedent – Singapore required their SafeEntry QR codes to be displayed at all businesses in May.

This announcement caused three things to happen over the following week. Firstly, the private developers with the most prevalent QR codes agreed that consolidation was necessary, and advised their customers to switch to the government QR code. Secondly, businesses largely complied, with the number of government QR codes increasing four times over the subsequent two weeks (from approximately 87,000 to 324,000). Thirdly, the presence of the disease in the community in New Zealand and the accompanying lockdown elevated the seriousness of the situation, and more people began to scan the QR codes. A few private systems still exist, but they are now all compatible with the NZ COVID Tracer QR code at least, even if the data is stored in a separate system.

The number of daily scans shot up, from approximately 30,000 per day before the second wave to over two million per day. The number of registered users also increased, from 640,000 before the second wave to just over two million users (approximately 50% of the adult population, although duplicate registrations are not accounted for) as of September 4. While the change in context was a significant driver for shifting user behaviours, moving away from the fragmented system clearly helped increase participation in the system too.

Unfortunately, this increase in participation came too late to be of significant help for the second wave. In the event of an outbreak, contact tracers need at least 14 days of movement logs for infected cases in order to help find close contacts. While the government did publish six exposure notifications through the app, and some close contacts were found faster because they had updated their contact details, it seems that ultimately the app is yet to find significant numbers of close contacts or new cases. As New Zealand now loosens movement restrictions, we can only hope that the current level of participation continues to grow for NZ COVID Tracer, and is sustained long enough to help defend against a potential third wave of cases in New Zealand.

**Leadership and Communication**

When a global pandemic catches the world by surprise, a strong response is needed to contain, mitigate, and recover from its impacts. In the case of contact tracing registers in New Zealand, many individuals wanted to do something to contribute, and the government opted to let the market sort it out. While people are to be lauded for utilizing their skills to support the broader community, the lack of co-ordination led to confusion, duplication of effort, and ultimately harmed the overall health response. This is not the fault of the individuals – the responsibility lies with the public health agencies to make good use of the available resources and to clearly communicate about what is and isn't needed. Competition between private entities in the open market might normally be desirable to drive innovation, but in a pandemic, we really need something that just works and supports progress on public health outcomes.

The context for the country was also highly important. Proclaiming that the pandemic was eliminated in New Zealand led the population to relax and become complacent – as no threat was perceived, actively scanning QR codes seemed unnecessary. Finding a way to communicate ongoing risk while at the same time celebrating success and projecting optimism is extremely challenging. In a country that has had strong communication with the public during the pandemic overall, the confusion around contact tracing registers has been an unfortunate blemish for New Zealand. This case study shows that fragmentation can lead to disparate and negative user experiences, which can harm trust in the system and lead to low participation. In the pandemic context, trust is one of the things we need the most for an effective response.

**Acknowledgements**

The author thanks the local Twitter community for contributing their images of QR codes to the dataset, and for engaging on digital contact tracing during the COVID-19 pandemic. The author also acknowledges members of Koi Tū: The Centre for Informed Futures for discussions around the use of technology for contact tracing.

***Dr. Andrew Tzer-Yeu Chen*** *is a Research Fellow with Koi Tū: The Centre for Informed Futures, a transdisciplinary think-tank at the University of Auckland, New Zealand. His background is in computer engineering, investigating computer vision surveillance and privacy. His research interests now sit at the intersection of digital technologies and society.*